\documentstyle{neu}
\newcommand{\be}{\begin{equation}}
\newcommand{\ee}{\end{equation}}
\newcommand{\Msun}{\mbox{$M_{\odot}\;$}}

\begin{document}

\title{THERMAL EVOLUTION OF ISOLATED NEUTRON STARS}

\author{Dany Page \\
        {\it Instituto de Astronom\'{\i}a,
          Universidad Nacional Aut\'onoma de M\'exico,
          M\'exico, D.F. 04510, MEXICO; page@astroscu.unam.mx}}

\maketitle

\section*{Abstract}

After a brief summary of neutron star cooling theory I present results which
emphasize the importance of baryon pairing in the neutron star core. 
I show how the thermal evolution may be totally controlled by pairing for
models which include only nucleons and models containing also hyperons.
Finally, I consider the thermal evolution of ultramagnetized neutron stars
whose existence has been inferred from the soft gamma repeaters.

\section{Introduction and Prelude}

The study of the cooling of neutron stars is being pursued principally
with the hope to constrain the structure of dense matter since 
young neutron stars cool by neutrino emission which comes mainly from
the matter at supranuclear densities in the deep interior.
The initial scenario, known as `The Standard Model', was based on the
modified Urca process.
This inefficient process leads to a slow cooling and the lack of detection of 
neutron stars, and their surface thermal emission, in several supernova 
remnants during the '60 and '70 stimulated the search for stronger neutrino
emission channels.
In 1965  Bahcall \& Wolf had shown that free pions, if present, would
enormously increase the neutrino emissivity:
a realistic mechanism for the presence of pions, pion condensation, was 
soon proposed and with it the dichotomy between `standard' and `exotic'
cooling scenarios (pions have nothing exotic, of course, but matter with
a pion condensate is exotic).
With the years, other possible fast cooling mechanisms have been 
proposed: quark matter, kaon condensation, the direct Urca process with 
nucleons and finally the direct Urca process with hyperons
(see Pethick 1992 for a review).

Naively, the standard model, on one side, and the fast cooling scenarios,
on the other, lead to very different predictions and there was great hope 
to be able to distinguish them by comparison of cooling models with
data: fast cooling scenarios predict much lower surface temperatures
at ages below $\sim 10^{5 - 6}$ yrs.
In this way, a cold (say $T_e << 10^6$ K) young (say $t < 10^5$ yrs)
neutron star would require fast neutrino cooling while a warm one
would be evidence in favor of the standard model.
However, things are more complicated than just the raw neutrino emission rate.
Baryon pairing can suppress the neutrino emission when the temperature $T$
is below the critical temperature $T_c$: a high $T_c$ can lead to such
strong suppression that, even with the most efficient neutrino emission
mechanism (the direct Urca process) the resulting surface temperature
can be higher than predicted within the standard model 
(Page \& Applegate 1992).
Unfortunately, the uncertainty on the value of $T_c$ in the core is
such that almost any surface temperature could be accommodated within almost
any fast cooling scenario.

With the launch of ROSAT came the confirmation of the detection of surface
thermal emission from several neutron stars (\"Ogelman 1995): 
the estimated surface temperatures lay close to the predictions of the
standard model and made some theorists (particularly myself) quite unhappy.
Later, a new and arguably more realistic fit of the Vela pulsar spectrum 
nevertheless gave a temperature estimate twice lower than the original one,
which seemed to seriously indicate the occurrence of some fast cooling agent 
at least in that neutron star (Page et al. 1996).
But it turned out that an essential process had been completely forgotten
in all cooling calculations:
neutrino emission due to the formation of Cooper pairs when the neutron
(or proton) liquid becomes superfluid (resp. superconducting).
This process is perfectly `standard' and, once taken into account, the 
standard model can be perfectly compatible with our present interpretation of
the data: {\em the standard model is alive and well alive}
(see Schaab et al 1997; Page 1998; Yakovlev {\em et al.} 1998).
Non standard scenarios are of course also compatible with the data.

I have presented earlier (Page 1998) a less sketchy discussion of the above
lines, to which I refer the reader for more details and references.
In the next sections I consider some complementary issues.

\begin{figure}
   \vspace{7.5cm}
\includegraphics{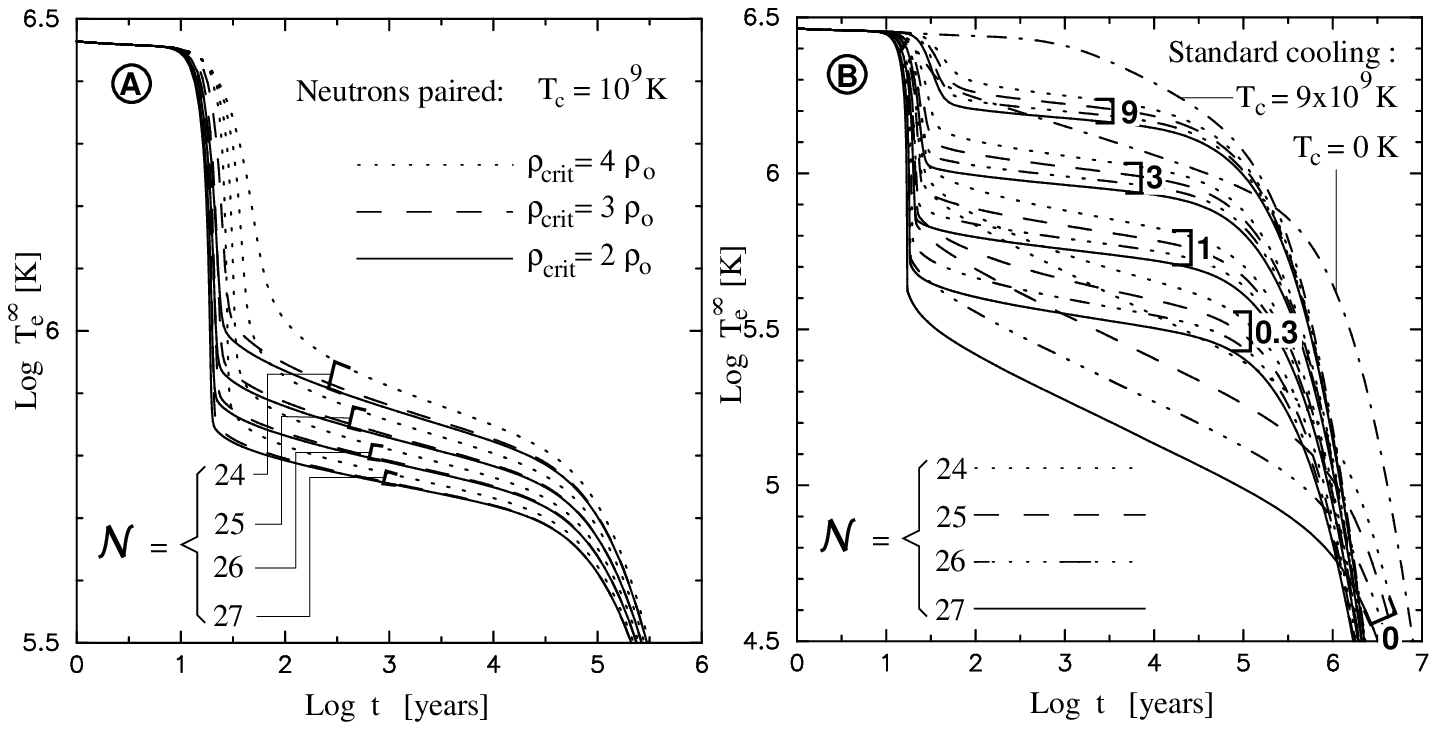}
\caption{{\bf Neutron stars as thermometers for nucleon pairing:} see text. 
         In B, the numbers on the curves indicate the assumed uniform value 
         of the 
         neutron $^3$P$_2$ pairing $T_c$, in units of 10$^9$ K, and two 
         standard
         cooling curves are also shown for comparison. (From Page 1995).
         \label{fig1}}
\end{figure}

\section{`Measuring' the baryon pairing critical temperature}

I want first to illustrate the extreme sensitivity of the fast neutrino cooling
to the baryon pairing critical temperature, $T_c$, by showing some previous 
(but little known) results (Page 1995).
It is even more than just sensitivity: if pairing occurs in the whole core 
then the only thing which really matters is the value~of~$T_c$~.
The reason is simple: once $T$ reaches $T_c$ the neutrino emission is 
suppressed and the main difference between the various fast mechanisms is that
$T_c$ will be reached very very early, or just very early, i.e., a few seconds
of a few days after the birth of the neutron star.
When looking at the star thousands of years later this early phase does not 
make much of a difference, only the value of $T_c$ does. 
To illustrate this, I summarize the various fast neutrino emissivities by
\be
\epsilon_{\nu} = 
  10^{\cal N} \cdot T_9^6 \;\;\; {\rm erg \; cm^{-3} \; sec^{-1}}
\ee
where {$\cal N$} can take values from 24, similar to the effect of a kaon 
condensate, up to 27, corresponding to the nucleon direct Urca process 
($T_9 \equiv T/10^9$ K).
Another parameter which has only a limited relevance is the total mass of 
the inner core (which I call the `pit') where this fast neutrino emission is 
taking place:
it depends on the equation of state (EOS) and the critical density 
$\rho_{crit}$ for the onset of fast emission.
Figure~1 shows the results:
$T_c$ is assumed uniform over the whole core for simplicity and in [A] 
I vary both $\rho_{crit}$ and ${\cal N}$ while in [B] $T_c$ is also varied 
(but not $\rho_{crit}$ since its effect is so small, as shown in [A]).
It is clear that, for a given value of $T_c$, changing ${\cal N}$ and 
$\rho_{crit}$ has little effect but changing $T_c$ can span a large range of
surface temperatures at ages between $\sim$ 10$^2$ to 10$^5$ years.
All the curves of Figure~1 correspond to a fixed mass of 1.4 \Msun and fixed 
EOS (Friedman \& Pandharipande 1981): 
changing the star mass and/or the EOS changes the mass of the pit which is 
equivalent to changing $\rho_{crit}$ and has thus very little effect.
The small differences between different values of ${\cal N}$ come from the 
fact that the neutrino emission is not instantaneously turned off below $T_c$
but suppressed and for a higher ${\cal N}$ the suppression takes a longer time
to complete.

One could thus consider fast cooling neutron stars as ``thermometers for the 
highest 
$T_c$ superfluid in the Universe'' by comparing results of Figure~1B
with estimated surface temperatures (see, e.g., Figure~2B):
{\bf T$_c \sim$ 2~-~3~$\times$~10$^9$~K}.
Since $T_c$ is density dependent this is its lowest value in the pit.
If several fast emission processes are possible, then it its the lowest value
for all different paired components (see, e.g., next section).
These results may be seriously affected by the occurrence of internal heating 
and/or the presence of light elements at the surface (Page 1997, 1998):
both of these may be resolved observationally and hopefully the future will 
tell us if they are, or are not, important.

\section{`Neutron Stars' with Hyperons}

\begin{figure}[t]
   \vspace{8cm}
\includegraphics{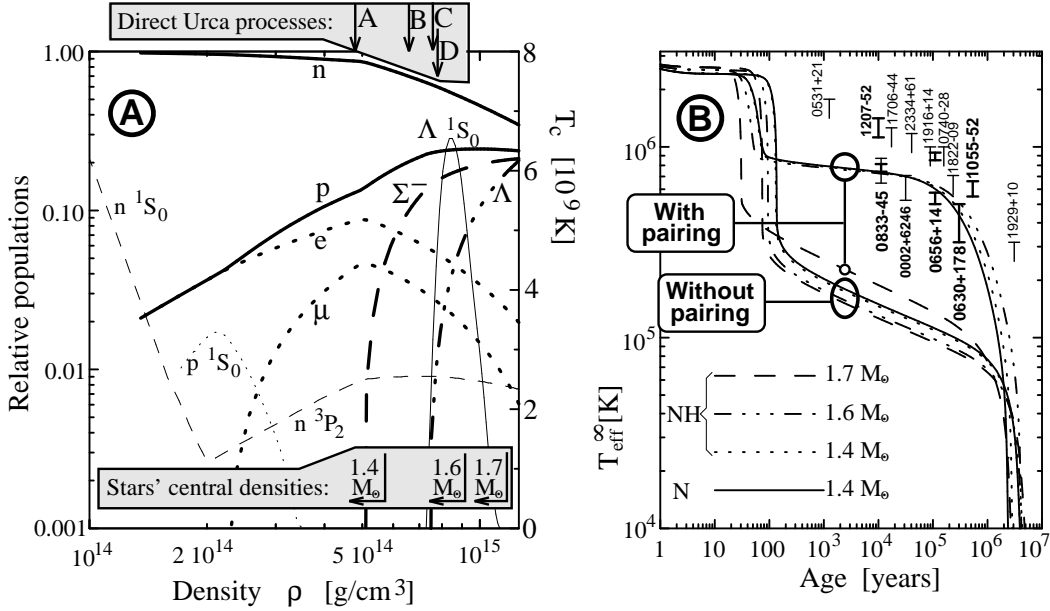}
\caption{{\bf Cooling of neutron stars containing hyperons}.
         The left panel shows the chemical composition of a model
         of dense matter with nucleons and hyperons 
         (after Glendenning \& Moszkowski 1991) and the pairing
         critical temperatures I assume for neutrons, protons
         and $\Lambda$s.
         The right panel shows the cooling of three stars containing
         nucleons and hyperons (Model NH) and only nucleons (Model N)
         for both cases with or without pairing.
         For reference to the data see Page (1997) and Zavlin {\em et al.}
         (1998) for 1E 1207-52:
         in short, the bigger the label the more reliable the temperature
         estimate.
         \label{fig2}
}
\end{figure}

There is presently a growing interest in the properties of neutron stars 
containing hyperons in their inner core.
I consider them here both for their intrinsic interest and to illustrate further
the considerations of the previous section.
Some features of the particular models I use are described in Figure~2A.
Four different direct Urca processes are possible in this case:
\begin{center}
\begin{tabular}{lrcl}
A \makebox[1in] & $n        \rightarrow p       + e^- + \overline{\nu}$ & and
                & $p       + e^- \rightarrow n        + \nu $ \\
B \makebox[1in] & $\Sigma^- \rightarrow n       + e^- + \overline{\nu}$ & and
                & $n       + e^- \rightarrow \Sigma^- + \nu $ \\
C \makebox[1in] & $\Sigma^- \rightarrow \Lambda + e^- + \overline{\nu}$ & and
                & $\Lambda + e^- \rightarrow \Sigma^- + \nu $ \\
D \makebox[1in] & $\Lambda  \rightarrow p       + e^- + \overline{\nu}$ & and
                & $p       + e^- \rightarrow \Lambda  + \nu $ \\
\end{tabular}
\end{center}
plus the similar processes with electrons replaced by muons.
The threshold densities are indicated in Figure~2A.
The cooling curves of Figure~2B show that when no pairing is taken
into account the evolution of stars with or without hyperons are similar.
This is simply because all these stars have process A allowed, which is the
most efficient of all, and B, C, and D can only act as small additions to A.
The surface temperatures obtained are obviously much below all current
observational values.
All these models have an envelope made of iron and no heating mechanisms are 
included:
an accreted envelope and strong heating could reduce the discrepancy between
models and observation (Page 1997, 1998).

More realistic models must include baryon pairing.
The calculation of the critical temperature $T_c$ for the various possible
pairings is a very tricky task, and one which has so far been barely touched
in presence of hyperons.
The values of $T_c$ I assume are plotted in Figure~2A and the two most 
important ones here are for neutron $^3$P$_2$ and $\Lambda$ $^1$S$_0$ pairings:
they control the cooling.
The neutron $^3$P$_2$ $T_c$ is taken such that the model without
hyperons (continuous curve) `fits' several observed temperatures
($T_c \sim 2 - 3 \cdot 10^9$ K).
From the four direct Urca processes, A and B are thus suppressed when the star
cools below the neutron $T_c$ and as a result the 1.4 \Msun star with hyperons 
(only $\Sigma^-$'s in this case) has the same cooling history as the one
without hyperons.
For the two stars of masses 1.6 and 1.7 \Msun, processes C and D will not be 
suppressed by neutron pairing and require hyperon pairing.
Recently, Balberg \& Barnea (1997) performed the first calculation of $\Lambda$
$^1$S$_0$ pairing in dense matter and obtained the values I plot in Figure~2A.
As all such calculations, this one should be taken with extreme care but it
leads to some interesting results.
The high $T_c$ for $\Lambda$ pairing suppresses so strongly processes C and D 
in  the 1.6 \Msun star that its evolution is again indistinguishable from the
previous cases.
Only for the most massive star is there a central region where C and D
can go uninhibited because the $\Lambda$'s $T_c$ vanishes.
This central region has a mass of only $\sim$~0.1~\Msun but the direct Urca
process is so efficient that the resulting evolution is similar to the cases
where there is no pairing at all.
We can thus have stars with very different cores, as the 1.4 \Msun star with
only nucleons and the 1.4 and 1.6 \Msun stars with hyperons, which have almost
identical cooling histories and stars with similar cores, as the 1.6 and 
1.7 \Msun stars with hyperons, which have very different histories:
it all comes down to pairing.

\section{Ultramagnetized Neutron Stars}

The possibility that the persistent X-ray emission from the three soft gamma
repeaters (SGR) is thermal emission from young ultramagnetized neutron stars
has raised much interest recently. 
These objects are characterized by X-ray luminosities which are much higher
than the ones of ordinary neutron stars.
The critical issue here is whether these luminosities 
can or cannot be explained simply by the cooling of the neutron star.
Every piece of physics which can raise the surface X-ray emission has to be
included and analyzed; 
and every piece of physics which can decrease it must be avoided !

On the positive side is the ultra strong magnetic field which significantly 
increases heat transport in the envelope
(Usov 1997; Heyl \& Hernquist 1997a, b: HHa \& HHb hereafter) and also
the possible presence of light elements in the envelope which has a similar and
cumulative effect (Chabrier {\em et al.} 1997; HHa).
A second mechanism is of course baryon pairing in the core with its suppressing
effect on the neutrino emission.
By combining these two ingredients we can obtain the kind of results shown in
Figure~3.
For iron envelopes and little pairing in the core my results are practically
identical to the ones of HHa (whose results of envelope models I boldly copied
from their figures) and with pairing the luminosity at the interesting ages
of a few thousand years is of course higher.
In the case of hydrogen envelopes and little pairing in the core I obtain
luminosities more than a factor two lower than HHa: this is due to the fact
that with such high fluxes in the envelope its temperature gradient extend to
higher densities, about $10^{12}$~gm~cm$^{-3}$, than the ones considered by 
HHa, who assumed isothermality at $\rho$ above $10^{10}$~gm~cm$^{-3}$.
Since I do not include the effect of the magnetic field on the thermal
conductivity at $\rho~>~10^{10}$~gm~cm$^{-3}$, this may not be reliable but it
certainly tells us that ultramagnetized envelope calculations should be 
extended to higher densities.

On the negative side would be any fast neutrino emission mechanism.
Controlling fast neutrino emission by baryon pairing would probably require 
unreallistically high values of $T_c$ to obtain such large surface temperatures.
One may state, tentatively, that {\em if} the soft X-ray emission of the SGRs
is from the cooling of the neutron star then this must be `standard cooling', 
i.e., these neutron stars do not have `exotic' matter in their core.
SGR 0526-66 is far too bright for its emission to be purely from the cooling
(as noted already by HHb) and some other mechanism(s) 
(see, e.g., Thompson \& Duncan 1996) must be at work.
The same mechanism(s) could also be at work in the other objects and if
efficient enough could be compatible with fast cooling scenarios, or, simply,
the observed X-rays have no, or little, relation with thermal emission.

\begin{figure}[t]
   \vspace{7.5cm}
\includegraphics{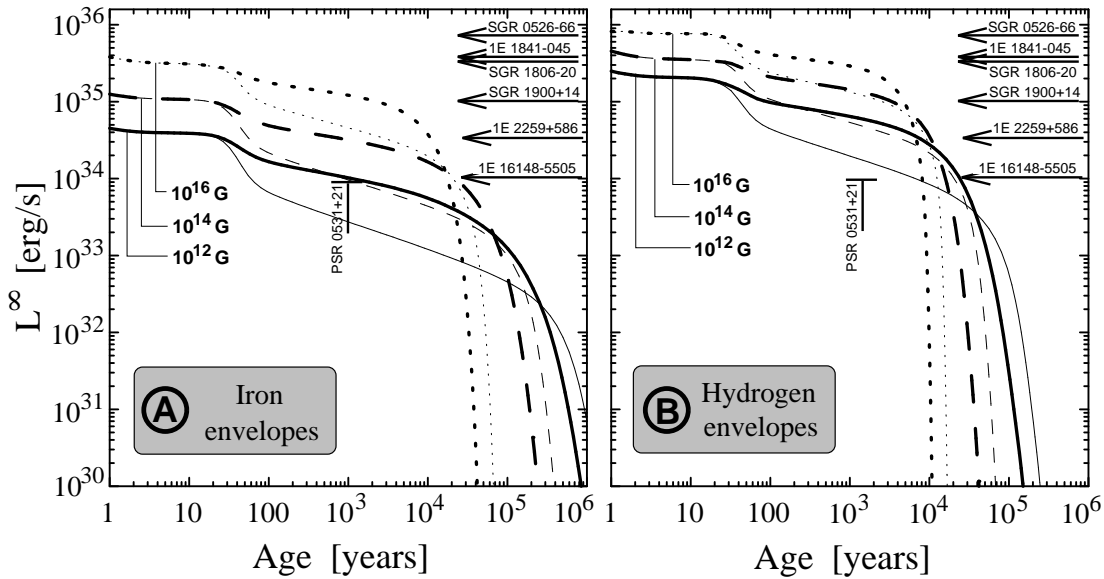}
\caption{{\bf Cooling of ultramagnetized neutron stars}.
         I consider the cases of iron and hydrogen envelopes separately:
         for a given interior temperature the hydrogen envelope implies
         an effective temperature about 70\% higher.
         Thin lines show models with most of the core neutrons unpaired
         while thick lines have a high neutron $^3$P$_2$ $T_c$ of the 
         order of $5 \times 10^9$ K.
         Data from:
         Rothschild {\em et al.} (1994) for SGR 0525-66, 
         Vasicht \& Gotthelf (1997) for 1E 1841-045,
         Murakami {\em et al.} (1994) for SGR 1806-20,
         Hurley {\em et al.} (1996) for SGR 1900+14,
         Parmar {\em et al.} (1997) for 1E 2259-586 and
         Gotthelf {\em et al.} (1997) for 1E 16148-5505.
\label{fig3}}
\end{figure}

\section{Postlude}

Faced with so many different models, and uncertainties, 
the naive expectation to learn about neutron star interiors by studying
only their thermal evolution is definitely doomed.
The thermal response of a neutron star to a glitch (see, e.g., Hirano {\em et al.}
1997) seems to be a promising complementary tool.
When combined with other tools, such as the maximum mass and rotational
frequency, dynamical studies of glitches, magnetic field evolution,
kHz QPO's, modeling of core collapse supernovae, detection of
neutrinos and gravitational waves, ...
plus laboratory results and more theoretical work, we may reach the Holy Grail
of understanding neutron star interiors and dense matter.
Future observational results form AXAF, XMM, Astro-E, and others,  will allow
us to resolve the all important issues (Page 1997) of
the chemical composition of the surface,
and the occurrence of heating mechanisms, among others.
The input of any strong constraint from these studies into cooling theories
would probably allow us to eliminate numerous models, correct unrealistic
assumptions and finally contribute to  `The Quest'.

\smallskip
\begin{footnotesize}
{\bf Acknowledgments:}
this work was supported by grants from UNAM-DGAPA (\#~IN 105495)
and Conacyt (\#~2127P-E9507). I also thank the organizers of the meeting
for their financial support during my stay in Tokyo.
\end{footnotesize}

\section{References}

\re
Bahcall, J. N., \& Wolf, R. W. 1965,
Phys. Rev., 140, B 1452

\re
Balberg, S., \& Barnea, N. 1997,
ECT* preprint (nucl-th/9709013)

\re
Chabrier, G., Potekhin, A.Y., \& Yakovlev, D. G. 1997,
ApJ, 477, L99

\re
Friedman, B., \& Pandharipande, V.R. 1981,
Nucl. Phys., A361, 502

\re
Glendenning, N. K., \& Moszkowski, S. A. 1991
Phys. Rev. Lett., 67, 2414

\re
Gotthelf, E. V., Petre, R., \& Hwang, U. 1997,
Ap. J. Lett., 487, L175

\re
Heyl, J. S., \& Hernquist, L. 1997a,
Ap. J. Lett., 489, L67 (HHa)

\re
Heyl, J. S., \& Hernquist, L. 1997b,
Ap. J. Lett., 491, L95 (HHb)

\re
Hirano, S., Shibazaki, N., Umeda, H., \& Nomoto, K. 1997,
Ap. J. 491, 286

\re
Hurley, K., {\em et al.} 1996,
Ap. J. Lett., 463, L13

\re
Murakami, T. {\em et al.} 1994,
Nature, 368, 127

\re 
\"Ogelman, H. 1995,
in {\em The Lives of the Neutron Stars}, 
Eds. M. A. Alpar, \"U. Kizilo\v{g}lu, \& J. van Paradijs
(Kluwer Academic Publishers: Dordrecht), 101

\re
Page, D. 1995,
Rev. Mex. F\'{\i}s., 41, Supl. 1, 178
(astro-ph/9501071)

\re
Page, D. 1997,
ApJ, 479, L43

\re
Page, D. 1998,
in {\em The Many Faces of Neutron Stars},
Eds. A. Alpar, R. Buccheri, \& J. van Paradijs
(Kluwer Academic Publishers: Dordrecht), in press.
(astro-ph/9706259)

\re
Page, D., \& Applegate, J.~H. 1992,
ApJ, 394, L17

\re
Page, D., Shibanov, Yu. A., \& Zavlin, V. E. 1996,
in {\em R\"ontgenstrahlung from the Universe},
Eds. H. U. Zimmermann, J. Tr\"umper, \& H. Yorke
(MPE Report 263: Garching), 173
(astro-ph/960187)

\re
Parmar, A. N., {\em et al.} 1997,
submitted to A\&A (astro-ph/9709248)

\re
Pethick, C. J. 1992,
Rev. Mod. Phys., 64, 1133

\re
Rothschild, R. E., Kulkarni, S. R., \& Lingenfelter, R. E. 1994,
Nature, 368, 432

\re
Schaab, Ch., Voskresensky, D., Sedrakian, A. D., Weber, F., \& Weigel, M. K. 1997,
A\&A, 321, 591

\re
Thompson, Ch., \& Duncan, R. 1996,
Ap. J., 473, 322

\re
Usov, V. V. 1997,
A. \& A. Lett., 317, L87

\re
Vasicht, G., \& Gotthelf, E. V. 1997,
Ap. J. Lett., 486, L129

\re
Yakovlev, D. G., Kaminker, A. D., \& Levenfish, K. P. 1998,
{\em these proceedings}

\re
Zavlin, V. E., Pavlov, G. G., \& Tr\"umper, J. 1998,
{\em these proceedings}

\end{document}